\documentclass[final,3p,times,twocolumn]{elsarticle}
\usepackage{graphicx}
\usepackage{amssymb}
\usepackage{amsmath}
\usepackage{hyperref}
\usepackage{gensymb}

\usepackage{amsmath}
\usepackage{makeidx}
 
 \journal{Physica D}

\usepackage[usenames]{color}
\usepackage[normalem]{ulem}
\usepackage{gensymb}

\usepackage{hyperref}
\usepackage[T1]{fontenc} 
\begin{document}

\title{A symbiotic  nondipolar  droplet  supersolid  in a binary dipolar-nondipolar Dy-Rb mixture}


 \author[int]{S. K. Adhikari}
\ead{sk.adhikari@unesp.br}

\address[int]{Instituto de F\'{\i}sica Te\'orica, UNESP - Universidade Estadual Paulista, 01.140-070 S\~ao Paulo, S\~ao Paulo, Brazil}

\begin{abstract}

We demonstrate the formation of two types of symbiotic  nondipolar  droplet supersolid   in a binary dipolar-nondipolar mixture with an  interspecies atraction, where the dipolar (nondipolar)  atoms are trapped (untrapped).   In the absence of an  interspecies attraction, in  the first type, a dipolar droplet supersolid exists,  whereas in the second type, there are no droplets in the dipolar component.  To illustrate, we consider  a  $^{164}$Dy-$^{87}$Rb mixture, where the  {\it untrapped}  $^{87}$Rb supersolid sticks to the trapped $^{164}$Dy supersolid due to the interspecies attraction and forms a symbiotic    supersolid with overlapping droplets. The first (second) type of symbiotic supersolid emerges for the  scattering length  $ a_1=85a_0$ ($a_1=95a_0$) of $^{164}$Dy atom, while under  an appropriate trap a dipolar droplet supersolid exists (does not exist) for no interspecies interaction, where $a_0$ the Bohr radius.  This study is based on  the numerical solution of an improved binary mean-field model, where we introduce an intraspecies   Lee-Huang-Yang interaction in the  dipolar   component, which stops a dipolar collapse and  forms  a  dipolar supersolid.To observe this symbiotic droplet supersolid, one should prepare the corresponding fully trapped dipolar-nondipolar supersolid   and then remove the trap on the nondipolar atoms.

\end{abstract}


\maketitle

\section{Introduction}

The observation of a dipolar Bose-Einstein condensate (BEC)
 of $^{52}$Cr 
 \cite{crrev,cr,cr1,saddle,ExpCr,52Cr}, 
  $^{166}$Er \cite{Er16},  
  $^{168}$Er \cite{ExpEr},
  $^{164}$Dy \cite{dy,dy1,dy2,ExpDy}   atoms
with large magnetic dipole
moments ($6\mu_{\mathrm{B}}, 7\mu_{\mathrm{B}}$, and  $10\mu_{\mathrm{B}}$, respectively, for chromium, erbium, and dysprosium atoms,  where $\mu_{\mathrm{B}}$ is a Bohr magneton)  initiated  intense  research  activity  to find any new physics emerging from 
the anisotropic long-range dipolar interaction,   in contrast to the isotropic zero-range contact interaction  
of a  more common dilute 
BEC of alkali-metal atoms with negligible dipole moment. The most remarkable phenomenon  in a strongly dipolar harmonically trapped BEC is the observation of a single droplet \cite{drop1,drop2} of size much smaller than the harmonic oscillator trap length.    With the increase of the number of atoms in the dipolar BEC, multiple droplets arranged on a triangular  or a linear  lattice  was observed in a strongly dipolar BEC of $^{164}$Dy \cite{Dy11,Dy12}, $^{162}$Dy \cite{Dy13,Dy14,Dy15}, and $^{166}$Er \cite{Dy11,Dy12} atoms and 
 supersolid properties for the linear arrangement of droplets was established.  The supersolid property for the triangular arrangement of droplets was also established  later \cite{tri21}.
 A supersolid \cite{1,2,3,4,5,6}, or a solid superfluid, is a quantum
state of matter simultaneously possessing the properties
of both a solid and a superfluid,  where a spontaneous density modulation and a global phase coherence coexist. A supersolid has
a spatially-periodic crystalline structure as a solid, breaking continuous translational invariance, and also enjoys
frictionless flow like a superfluid, breaking continuous
gauge invariance.  In a trapped quasi-two-dimensional (quasi-2D)
dipolar BEC, the formation of honeycomb lattice, stripe,
square lattice and other periodic patterns in density, were
also found  \cite{34,35,39,40,41,th3} in theoretical studies.

It seems that a  quasi-2D dipolar BEC supersolid can only exist under  a confining box \cite{40} or harmonic \cite{drop1,drop2}  trap. It will be desirable 
to create a BEC supersolid  without a confining trap. We demonstrate that two types of  untrapped 
{\it nondipolar} BEC supersolid can be created in a binary dipolar-nondipolar BEC  in the presence of a trapped dipolar 
supersolid. {FThere have been extensive studies of multidimensional self-trapping of matter 
\cite{malo,miha} without a trap.}
 In the absence of an interspecies
interaction, in the first type, a dipolar droplet-lattice  supersolid exists, whereas in the second type, there
are no droplets in the dipolar component. In both types, untrapped nondipolar atoms are not localized in absence of an interspecies attraction. 
In the first type, the intraspecies scattering length of dipolar atoms $a_1$  is kept sufficiently small, so that the dipolar attraction dominates over the contact repulsion and allows the formation of a dipolar supersolid in absence of an interspecies attraction.   In the second type, a larger value of $a_1$ makes the contact repulsion  sufficiently strong so that   there is no dipolar droplet formation without an interspecies attraction.

The shape 
of such a stable  untrapped nondipolar BEC supersolid  bound 
 in a trapped dipolar BEC supersolid  in the form of a symbiotic dipolar-nondipolar supersolid 
 is controlled by the
interspecies attraction,
whereas that of the trapped dipolar BEC supersolid  is determined by the 
underlying trap. 
This nondipolar BEC supersolid  will be termed quasi-free, as, being untrapped,  it  
can easily oscillate and move inside
the  trapped dipolar  BEC  supersolid  responsible for its binding.   
In a previous study we demonstrated the formation of a quasi-free untrapped dipolar droplet bound in a harmonically trapped nondipolar BEC \cite{2013}.
In this study we consider a binary dysprosium-rubidium  $^{164}$Dy-$^{87}$Rb 
mixture, with a smaller number of  $^{87}$Rb atoms compared to that of $^{164}$Dy atoms, 
and consider binary supersolids of both triangular- and square-lattice  symmetry.  The lattice constant and lattice structure of the dipolar and nondipolar  
supersolids are identical with the same number  of overlapping droplets in both components.

A dipolar mean-field Gross-Pitaevskii (GP)  equation cannot account for  the formation of a dipolar droplet or a droplet-lattice supersolid in a trapped  strongly dipolar BEC  due to a collapse instability resulting from the strong dipolar attraction \cite{52Cr,collapse}, when the dipolar attraction dominates over the contact repulsion.  
 An improved mean-field model including a higher-order perturbative  Lee-Huang-Yang (LHY) type \cite{lhy} repulsive interaction, appropriately modified due to the strong dipolar interaction \cite{qf1,qf3,qf2,sepa}, can stop the collapse of the strongly dipolar BEC and 
allow the formation of a dipolar droplet or a dipolar droplet-lattice  supersolid. 
{ We base the present study on the numerical solution of this improved binary dipolar-nondipolar mean-field model for the dipolar-nondipolar $^{164}$Dy-$^{87}$Rb mixture, where we introduce the LHY interaction only in  the Hamiltonian of the dipolar component, which suffers from the collapse \cite{collapse} instability for a large number of dipolar atoms confined in a quasi-2D trap.  As there will be no collapse in  the nondipolar component,  containing a much smaller number of atoms than that in the dipolar component, the effect of the LHY interaction in this component will be negligibly  small and  will not  be included in this study. 
  In another recent study \cite{halder} on  induced supersolidity in a $^{164}$Dy-$^{166}$Er mixture,  the LHY interaction was introduced only in the strongly dipolar $^{164}$Dy component, as the $^{166}$Er component was not in a
dipole-dominated interaction regime and was stable against
the mean-field collapse.} 
   The usage of a small number of atoms in the nondipolar component  eliminates the background atom cloud to a minimum in the  nondipolar supersolid
 thus generating a clean nondipolar supersolid.  However, if the number of nondipolar atoms is too small a large  number of droplets cannot be formed.

   The quasi-2D 
  trap has a stronger confinement  along the polarization $z$ direction and a circularly symmetric weaker confinement  in the transverse
   $x$-$y$  plane  with the same angular frequencies as used in some experimental  \cite{tri21,Ex2} and theoretical \cite{39,th1,Th2} investigations. To study the symbiotic supersolid of the first type, we take the dipolar scattering length 
   $a_1=85a_0$ with $a_0$ the Bohr radius, which enhances the dipolar attraction so as to form a dipolar supersolid \cite{39}.  
   To study the symbiotic supersolid of the second type, we take the dipolar scattering length 
   $a_1=95a_0$, which enhances the contact  repulsion  so as to exclude any dipolar droplet formation in absence  of an interspecies attraction.
   However, for an appropriate interspecies contact attraction, in both cases an overlapping symbiotic dipolar-nondipolar   $^{164}$Dy-$^{87}$Rb supersolid is formed.
   
 
We also demonstrate the possibility of observing the  symbiotic  dipolar-nondipolar supersolid in a laboratory. 
For this purpose we consider a binary dipolar-nondipolar $^{164}$Dy-$^{87}$Rb mixture where both the components are trapped. As the rubidium atoms are attached to the dysprosium atoms in the dipolar supersolid by the interspecies contact attraction, and not by the external trap,  the densities of the  
$^{164}$Dy-$^{87}$Rb mixture are essentially independent of whether the   $^{87}$Rb atoms are trapped or not. However, it is easier to create the  dipolar-nondipolar $^{164}$Dy-$^{87}$Rb supersolid mixture in a laboratory with a harmonic trap on both the components. Once the supersolid mixture is generated, the trap on the nondipolar atoms could be removed to generate the desired quasi-free nondipolar supersolid.

In Sec. 2  we present the  improved binary mean-field dipolar-nondipolar model  including the repulsive LHY interaction in the dipolar component, which we use for the study of the present symbiotic dipolar-nondipolar supersolid.  
The results of numerical calculation are illustrated in Sec. 3.  In addition to presenting the results of densities of a single-component dipolar BEC and of the binary  dipolar-nondipolar symbiotic supersolid, we also present results for the appearance of states with different number of droplets from an energetic 
consideration.   Starting from a fully trapped dipolar-nondipolar supersolid, we demonstrate the possibility of creating the present symbiotic dipolar-nondipolar supersolid by removing the trap from the nondipolar atoms.
Finally, in Sec. 4 we present a brief summary of our findings.

\section{Mean-field model for a symbiotic dipolar-nondipolar supersolid}

We consider a binary BEC, where one of the species is dipolar and the 
other nondipolar, interacting via  interspecies and intraspecies interactions.  The 
mass, number of atoms, magnetic dipole moment, and scattering length for the two species, denoted by  $ i=1$, 2,
are given by $m_i, N_i, 
\mu_i, a_i,$ respectively. The first   species ($i=1$) of atoms  ($^{164}$Dy) 
has large   dipole moment, and is trapped  and polarized along the axial $z$ direction.
On the other hand,  the second  species ($i=2$) of atoms ($^{87}$Rb) is untrapped,  has negligible dipole moment ($\mu_2 \ll \mu_1$) and will be  taken to be   nondipolar. 
The  intraspecies ($V_i$) and 
interspecies ($V_{12}$)
interactions 
for two atoms  at positions $\bf r$ and $\bf r'$ are taken as \cite{crrev,expt}
\begin{eqnarray}\label{intrapot} 
V_1({\bf R})&=&
\frac{\mu_0\mu_1^2}{4\pi}V_{\mathrm{dd}}({\mathbf R})+\frac{4\pi 
\hbar^2 a_1}{m_1}\delta({\bf R }),\\
V_2({\bf R})&=& 
\frac{4\pi 
\hbar^2 a_2}{m_2}\delta({\bf R }),\\ \label{interpot} 
V_{12}({\bf R})&=& 
\frac{2\pi \hbar^2 a_{12}}{m_R}\delta({\bf R}),\\ \label{dp}
V_{\mathrm{dd}}({\mathbf R})  &=& 
\frac{1-3\cos^2\theta}{{\mathbf R}^3},
     \end{eqnarray}
where $\bf R \equiv (r-r')$ is the position vector joining the two atoms, $\mu_0$ is the permeability of free space, 
$\theta$ is the angle made by the vector ${\bf R}$ with the polarization 
$z$ direction,   $m_R=m_1m_2/(m_1+m_2)$ is the reduced mass of the two species of 
atoms, and $a_{12}$  is the interspecies dipolar-nondipolar scattering length. 
To compare the dipolar and contact interactions, the intraspecies 
dipolar interaction  is  expressed in terms of the dipolar length
$a_{\mathrm{dd}}$, defined by $  a_{\mathrm{dd}}\equiv 
\mu_0\mu_1^2m_1/(12\pi \hbar ^2).$   The effect of the corresponding intraspecies contact interaction is quantized by the scattering length $a_1$. 
The dimensionless relative dipolar length, defined by, 
\begin{equation}
\varepsilon_{\mathrm{dd}}     \equiv \frac{a_{\mathrm{dd}}}{a_1}  
\end{equation}
gives
the strength of the dipolar interaction relative to  the contact interaction  
and is useful to study  many properties of a dipolar BEC.

The angular frequencies for the axially-symmetric quasi-2D harmonic trap on the first species of dipolar atoms ($^{164}$Dy)
along $x$, $y$ and $z$ directions are taken as 
$\omega_x=\omega_y\equiv \omega_\rho$   ($\boldsymbol \rho =\{x,y\},$  $\rho^2=x^2+y^2,$) and 
$\omega_z (\gg \omega_\rho)$, while the second species of nondipolar atoms ($^{87}$Rb)  is untrapped. 
 With intraspecies and  interspecies interactions (\ref{intrapot})-(\ref{interpot}), the  improved coupled   GP
equations for the binary dipolar-nondipolar  BEC mixture can be written as \cite{mfb,mfb1,mfb2} 
\begin{align}
\label{eq1}
{\mbox i} \hbar \frac{\partial \psi_1({\bf r},t)}{\partial t}  &=
{\Big [}  -\frac{\hbar^2}{2m_1}\nabla^2+ \frac{1}{2}m_1  
(\omega_\rho ^2\rho^2+\omega_z^2 z^2 )
\nonumber\\ &
+ \frac{4\pi \hbar^2}{m_1}{a}_1 N_1 \vert \psi_1({\bf r},t) \vert^2
+\frac{2\pi \hbar^2}{m_R} {a}_{12} N_2 \vert \psi_2({\bf r},t) \vert^2
\nonumber 
\\  
&+\frac{3\hbar^2}{m_1}a_{\mathrm{dd}} N_1 \int  V_{\mathrm{dd}} ({\mathbf R})\vert\psi_1({\mathbf r'},t)\vert^2 d{\mathbf r}' \nonumber \\
&+\frac{\gamma_{\mathrm{LHY}}\hbar^2}{m_1}N_1^{3/2}
|\psi_1({\mathbf r},t)|^3
\Big] 
 \psi_1({\bf r},t),
\\
{\mbox i} \hbar \frac{\partial \psi_2({\bf r},t)}{\partial t} &=
{\Big [}  -\frac{\hbar^2}{2m_2}\nabla^2
+ \frac{4\pi \hbar^2}{m_2}{a}_2 N_2 \vert \psi_2({\bf r},t)\vert^2 
\nonumber
\\  & 
+\frac{2\pi \hbar^2}{m_R} {a}_{12} N_1 \vert \psi_1({\bf r},t)|^2
{\Big ]}  \psi_2({\bf r},t),
\label{eq2}
\end{align}
where ${\mbox i}=\sqrt{-1}$. 
The    LHY interaction  coefficient  $\gamma_{\mathrm{LHY}}$ \cite{lhy} in  Eq. (\ref{eq1}) is given by \cite{qf1,qf3,qf2}
\begin{align}\label{qf}
\gamma_{\mathrm{LHY}}= \frac{128}{3}\sqrt{\pi a^5} Q_5(\varepsilon_{\mathrm{dd}}), 
\end{align}
where  the auxiliary function $ Q_5(\varepsilon_{\mathrm{dd}})$ includes the  
correction to the       LHY interaction due to the dipolar interaction   and is
 given by \cite{qf1,qf2}
\begin{align}
 Q_5(\varepsilon_{\mathrm{dd}})&=\ (1-\varepsilon_{\mathrm{dd}})^{5/2}  {_2F_1} \left(-\frac{5}{2},\frac{1}{2};\frac{3}{2};\frac{3\varepsilon_{\mathrm{dd}}}{\varepsilon_{\mathrm{dd}}-1}\right) , 
\end{align}
where $_2F_1$ is the hypergeometric function.  Using 
 an integral representation of the hypergeometric function \cite{arxiv} $_2F_1$  the auxiliary function $Q_5$ 
{ can be written as   \cite{blakie}
\begin{align}\label{exa}
Q_5(\varepsilon_{\mathrm{dd}}) &
 \equiv\int_0^1 du(1-\varepsilon_{\mathrm{dd}}+3\varepsilon_{\mathrm{dd}}u^2 )^{5/2},  \\
 &=\
\frac{(3\varepsilon_{\mathrm{dd}})^{5/2}}{48}   \Re \left[(8+26\eta+33\eta^2)\sqrt{1+\eta}\right.\nonumber\\
& + \left.
\ 15\eta^3 \mathrm{ln} \left( \frac{1+\sqrt{1+\eta}}{\sqrt{\eta}}\right)  \right], 
\end{align}
where $\Re$ is the real part and
\begin{align}
  \eta = \frac{1-\varepsilon_{\mathrm{dd}}}{3\varepsilon_{\mathrm{dd}}}.
\end{align} 
For $\epsilon_{\mathrm{dd}} = 0$ and 1, the expression (\ref{exa}) is indeterminate
and the following limiting
 values of $Q_5$ are to be used  
$Q_5 (0) = 1$ and $Q_5(1) = 3 \sqrt 3/2$ [40].
In this paper we use expression (\ref{exa}) for $Q_5(\varepsilon_{\mathrm{dd}})$.
  The perturbative result (\ref{qf}) is  valid for weakly dipolar atoms 
($\varepsilon_{\mathrm{dd}}<1$), while $Q_5(\varepsilon_{\mathrm{dd}})$ is real  \cite{qf1,qf2}. The dipolar supersolid appears only for   strongly dipolar atoms   ($\varepsilon_{\mathrm{dd}}>1$), while the quantity $\sqrt \eta$  in Eq. (\ref{exa}) is imaginary, and consequently, the auxiliary function 
$Q_5(\varepsilon_{\mathrm{dd}})$ is  complex. However,
for $^{164}$Dy atoms, 
 the imaginary part of the auxiliary function $Q_5$
 is negligible in  comparison to its real part    \cite{young},  for the value of scattering length $a_1$ used in this study,  and will be neglected as has been done elsewhere \cite{39,40,th1,expt}.}

 Equations  (\ref{eq1}) and (\ref{eq2})  can be written in the following dimensionless form if we scale lengths in units of 
 $l = \sqrt{\hbar/m_1\omega_z}$,
time in units of $\omega_z^{-1}$,    angular frequency in units of $\omega_z$,
energy in units of $\hbar \omega_z$ and density $|\psi_i|^2$   
in units of $l^{-3     }$ 
\cite{th1,mfb,mfb1,mfb2}
\begin{align}
\label{eq3}
{\mbox i}  \frac{\partial \psi_1({\bf r},t)}{\partial t}  &=
{\Big [}  -\frac{1}{2}\nabla^2+ \frac{1}{2}
(\omega_\rho ^2\rho^2+ z^2 )
+ g_1 \vert \psi_1({\bf r},t) \vert^2
\nonumber 
\\  
&+g_{12}\vert \psi_2({\bf r},t) \vert^2 
+ g_{\mathrm{dd}}      \int  V_{\mathrm{dd}} ({\mathbf R})\vert\psi_1({\mathbf r'},t)\vert^2 d{\mathbf r}' \nonumber \\
&+\gamma_{\mathrm{LHY}}N_1^{3/2}
|\psi_1({\mathbf r},t)|^3
\Big] 
 \psi_1({\bf r},t),
\\
{\mbox i}  \frac{\partial \psi_2({\bf r},t)}{\partial t} &=
{\Big [}  -\frac{m_{12}}{2}\nabla^2
+ g_2 \vert \psi_2({\bf r},t)\vert^2
\nonumber
\\  & 
+g_{21} \vert \psi_1({\bf r},t)|^2 
{\Big ]}  \psi_2({     \bf r},t),
\label{eq4}
\end{align}
where
$m_{12}={m_1}/{m_2},$
$g_1=4\pi a_1 N_1,$
$g_2= 4\pi a_2 N_2 m_{12},$
$g_{12}={2\pi m_1} a_{12} N_2/m_R,$
$g_{21}={2\pi m_1} a_{12} N_1/m_R,$
$g_{\mathrm{dd}}= 3N_1 a_{\mathrm{dd}}.$
 
For a stationary state, Eqs. (\ref{eq3}) and (\ref{eq4}) can be obtained from a minimization 
of the following energy functional  (total energy)
\begin{align}\label{energyx}
E&= \frac{1}{2}\int d{\bf r}\Big[N_1|\nabla \psi_1({\bf r})|^2 
+m_{12} N_2|\nabla \psi_2({\bf r})|^2
\nonumber \\
&+N_1(\omega_\rho^2\rho^2+ z^2)|\psi_1({\bf r})|^2+
\sum_i   
 N_ig_i|\psi_i({\bf r})|^4
\nonumber \\&
+ {N_1}g_{\text{dd}}\int d{\bf r}' V_{\text{dd}}({\bf R})|\psi_1({\bf r'})|^2   |\psi_1({\bf r})|^2 
\nonumber \\
&+\frac{4}{5}\gamma_{\mathrm{LHY}}N_1^{5/2} |\psi_1({\bf r})|^5\Big]
 + N_1g_{12}\int d{\bf r}|\psi_1({\bf r})|^2   |\psi_2({\bf r})|^2.
\end{align}
 Equations (\ref{eq3}) and (\ref{eq4}) follow from an extremization of this energy functional via the rule
 \begin{align}
 \mbox{i}\frac{\partial \psi_i}{\partial t}= \frac{\delta E}{\delta \psi_i^ *} .
 \end{align}

\section{Numerical Results}

To study the formation of a quasi-free nondipolar supersolid in a dipolar-nondipolar mixture, the partial differential GP equations (\ref{eq3}) and (\ref{eq4}) are solved, numerically, using C/FORTRAN programs \cite{49} or their openmultiprocessing versions \cite{52,53}, using the split-time-step Crank-Nicolson method by imaginary- and real-time propagation using a space step of 0.1 and a time step 0.001 and 0.0005, respectively \cite{Santos01,CPC,CPC1}. The method of imaginary-time propagation is used to find the stationary states and that of real-time propagation is used to study the dynamics. The solution obtained by imaginary-time propagation is the same as that obtained by a minimization of energy (\ref{energyx}).
It is difficult to evaluate numerically  the divergent $1/|{\bf R}|^3$ term in the dipolar potential (\ref{dp})
in configuration space.
The contribution of the nonlocal dipolar interaction integral in Eq. (\ref{eq3})  is calculated numerically in momentum 
space by a fast Fourier transformation routine  using a convolution theorem \cite{49,Santos01}.  The Fourier transformation of the dipolar interaction to momentum space  is analytically known \cite{49}.   In this way the problem is solved
in momentum space and the solution in configuration
space is obtained by taking another Fourier transformation.

\begin{figure}[!t]

\begin{center}

\includegraphics[width=\linewidth]{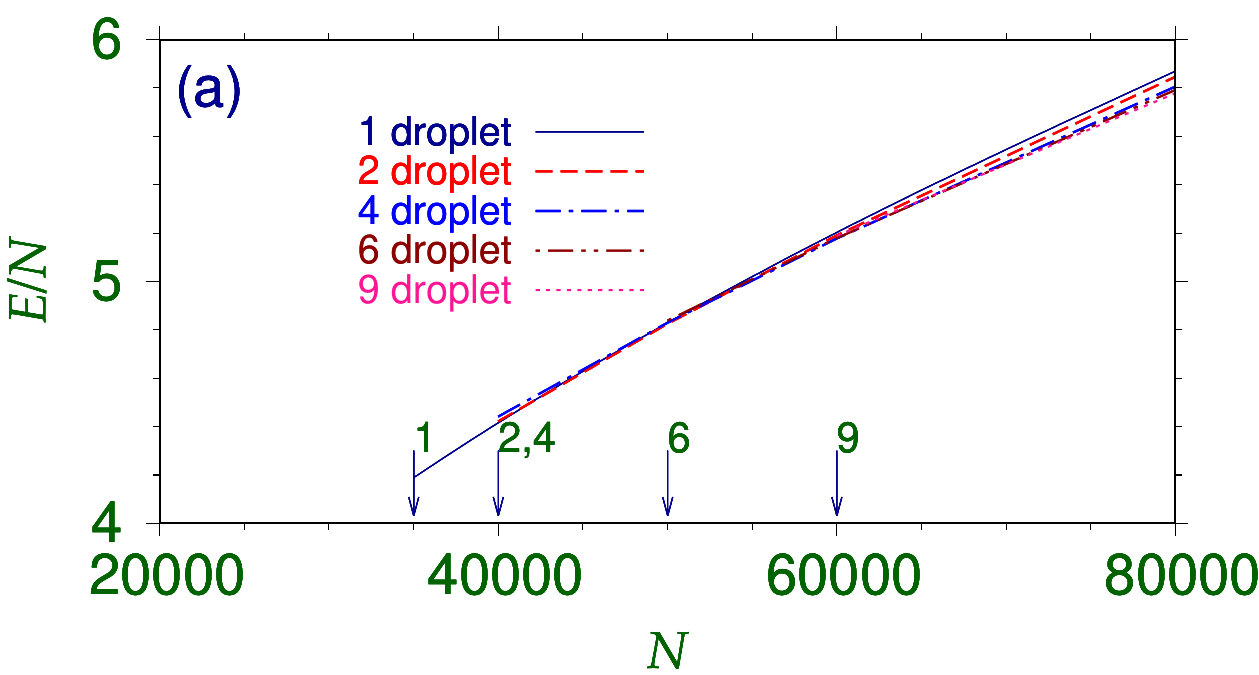}
\includegraphics[width=\linewidth]{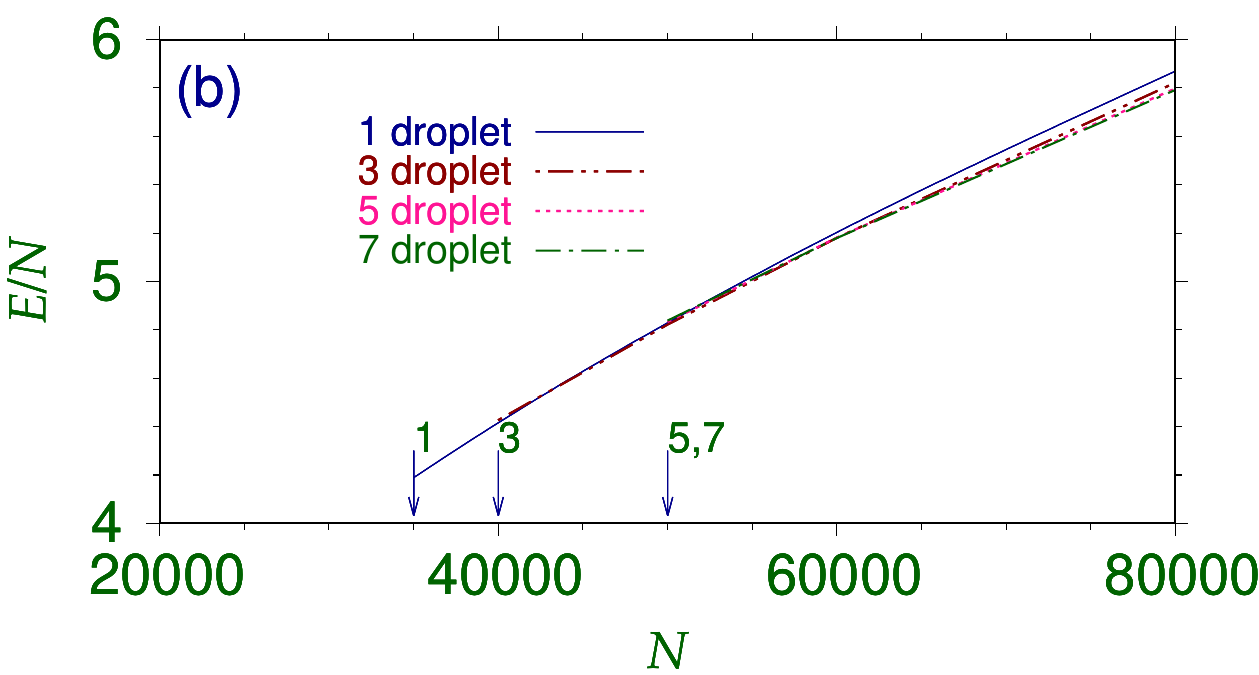}
  
\caption{ (Color online)
     Energy per atom $E/N$ versus the number of atoms $N$ of   spatially-symmetric states with a small number of droplets  in a dipolar BEC of $^{164}$Dy atoms.  The arrows indicate the threshold for the formation of a certain number of droplets. In Secs. 3.1 and 3.2 the scattering length of $^{164}$Dy atoms is
 $a=85a_0$.  
The   $^{164}$Dy atoms   are confined by  an  axially-symmetric quasi-2D trap with angular frequencies $\omega _z=2\pi \times 167$ Hz,  $\omega _ x=\omega _y=2\pi \times 33$ Hz and the dipolar length $a_{\mathrm{dd}}=130.8a_0$. All results  are  dimensionless with the scaling length $l=\sqrt{\hbar/m_1\omega_z}=0.607$ $\mu$m.     }
     \label{fig1}
\end{center}
\end{figure}

 The dipolar length of  $^{164}$Dy atoms  will be taken as  $a_{\mathrm{dd}}=130.8a_0$ \cite{ExpDy,rmp}.
Throughout this study
the dysprosium atoms are  confined by a quasi-2D  axially-symmetric trap   with angular  frequencies $\omega_\rho\equiv \omega_x=\omega_y=2\pi \times  33$ Hz, 
$\omega_z=2\pi \times 167$ Hz, as in some recent experiments on a quasi-2D hexagonal supersolid formation with $^{164}$Dy 
atoms
 \cite{tri21,Ex2} and also used in some theoretical investigations \cite{39,th1,Th2}, so that the length scale $l \equiv
\sqrt{\hbar/m_1\omega_z}=0.607$ $\mu$m. 
 The formation of the droplets is best confirmed by a consideration of the integrated reduced quasi-2D density 
 $n_i(x,y)$ defined by 
 \begin{equation}
 n_i(x,y)=\int_{-\infty}^{\infty}  dz|\psi_i(x,y,z)|^2.
 \end{equation}

\subsection{Single-component dipolar supersolid}

For the  dipolar  $^{164}$Dy atoms, we take the intraspecies scattering length as
$a_1\equiv a(^{164}$Dy)$ =85a_0$ for the study of a single-component dipolar supersolid and  the symbiotic supersolid of the first type, which is consistent with its experimental estimate   $a_1=92 (8)a_0$ \cite{tang}. A more recent update of this scattering length is $a_1=87 (8)a_0$  \cite{Dy11}.
A slightly smaller value of this scattering length makes the system more attractive and  facilitates the formation of dipolar droplets and a droplet-lattice  supersolid.
Other theoretical studies also used \cite{tri21,th1} a scattering length ($a_1=88a_0$) close to the  more recent estimate. The reason behind the use of a smaller $a_1$ is that no droplet-lattice supersolid could be obtained with the value $a_1=92a_0$ in theoretical models with the present trap.  The value $a_1=85a_0$ is within the limits  set by experiments \cite{Dy11,tang}.

 In Figs. \ref{fig1}(a)-(b) we illustrate the formation of different states with a  small number of droplets in a single-component trapped 
BEC of dipolar $^{164}$Dy atoms through a plot of energy per atom $E/N$  versus the number of atoms $N$.  We consider  only the   stable  spatially-symmetric  state of a fixed number (one to nine)  of droplets. 
For example, the three-droplet (four-droplet, five-droplet, seven-droplet, nine-droplet)  state is the one with the droplets forming an equilateral triangle (square, pentagon, hexagon, square).  More and more droplets are formed as the number of atoms increases.
 In these plots the arrows indicate  the threshold for the appearance of the states with a certain number of droplets.

\begin{figure}[!t]

\begin{center}

\includegraphics[width=\linewidth]{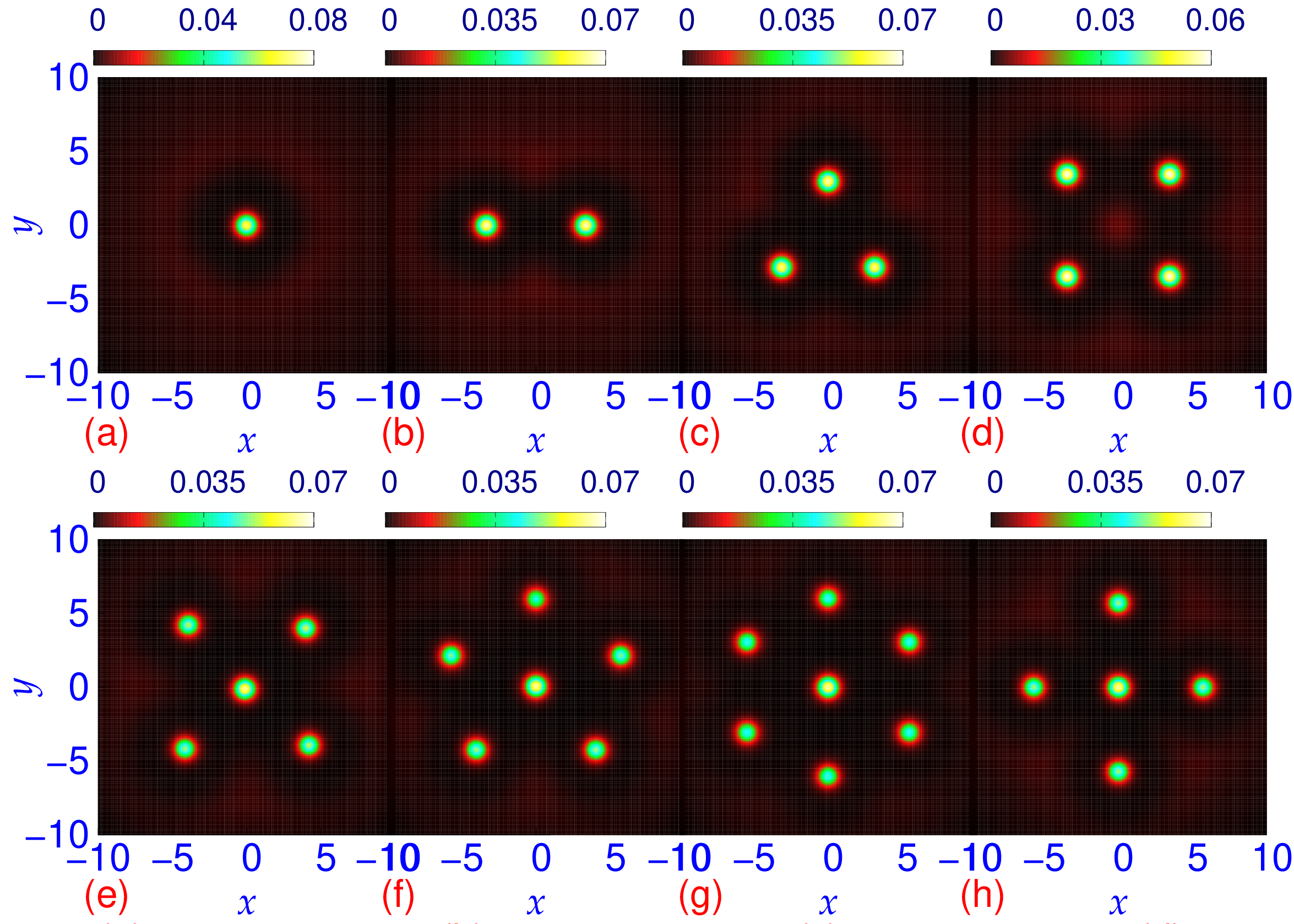}

\caption{ (Color online) Contour plot of quasi-2D density  $n(x,y)$ of spatially-symmetric states with 
(a)  one, (b)  two, (c) three, (d) four, (e) five, (f)  six, (g)  seven, and (h) five droplets, respectively,     in a single-component 
dipolar  $^{164}$Dy BEC of 50000 atoms.  The initial states in imaginary-time propagation  are appropriately placed 
states with one, two, three, four, five, six, seven, and nine,  droplets, respectively.  
  }\label{fig2}
\end{center}
\end{figure}

\begin{figure}[!t]

\begin{center}

\includegraphics[width=\linewidth]{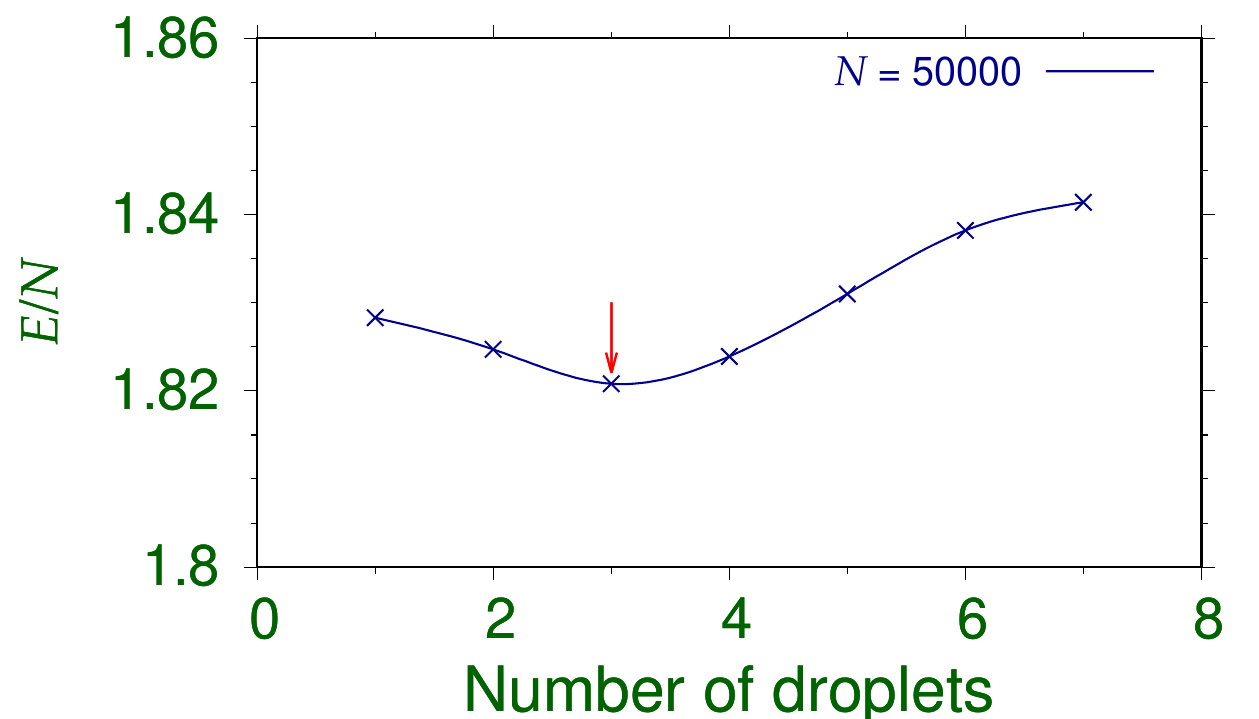}

\caption{ (Color online) Energy per atom $E/N$ of a $^{164}$Dy BEC of atoms, for different number of spatially-symmetric 
droplets illustrated in Fig. \ref{fig2}. The points are the result of calculation and the lines indicate the trend of energy evolution.    The arrow indicates the minimum of energy.  
  }\label{fig3}
\end{center}
\end{figure}

 Here we study the possibility of generating stable  droplet-lattice states  with different number of droplets  for a fixed number of dipolar atoms, as we will study in the following the possibility of generating symbiotic droplet-lattice supersolids with different number of droplets for a fixed number  of dipolar and nondipolar atoms. As there are many variables in  the binary dipolar-nondipolar mixture, we prefer to keep the number of atoms fixed in this study.
 The spatially-symmetric states considered in Fig. \ref{fig1} are illustrated explicitly in 
 Fig. \ref{fig2} through a contour plot of reduced quasi-2D density $n(x,y)$ of these states  with (a) one,
 (b) two, (c) three, (d) four, (e) five, (f) six, (g) seven, and (h) five,  droplets  for 
$N=50000$   $^{164}$Dy atoms. 
 The initial state in imaginary-time propagation in these calculations was a spatially-symmetric state  with (a)-(g) 1-7 and (h) 9 droplets, respectively, as explained in detail in Ref. 
\cite{39}. In Figs. \ref{fig2}(a)-(g)
the final state has the same number of droplets as  in the initial state. 
However, in Fig. \ref{fig2}(h) the initial state was a nine-droplet state on a square lattice,
which resulted in a five-droplet state.  Fifty thousand atoms is below  the threshold for the formation of a nine-droplet state as illustrated in Fig. \ref{fig1}. A nine-droplet state results as the number of atoms is increased to sixty thousand (not illustrated  here).  The states in Fig. \ref{fig2} are quasi-degenerate and also
have a background atom cloud, which reduces as the number of droplets increases for a fixed number of atoms. For example, the seven-droplet state in (g) has less background atom cloud than the one- or two-droplet state in \ref{fig2}(a) or \ref{fig2}(b), respectively.  The background atom cloud has contributed to the formation of droplets in the seven-droplet state with a cleaner background. In Fig. \ref{fig3}, we plot the energy of the different droplet-lattice states displayed in Fig. \ref{fig2} versus the number of droplets, which has a minimum for three droplets. This indicates that for $N=50000$ atoms 
the state with three droplets of Fig. \ref{fig2}(c) is the most probable one from an energetic point of view.

\subsection{Symbiotic dipolar-nondipolar supersolid of the first type}

In this study on the binary dipolar-nondipolar $^{164}$Dy-$^{87}$Rb  mixture,
 we fix the intraspecies scattering length  of  $^{87}$Rb atoms at its experimental value       $a_2\equiv a(^{87}$Rb$)=109a_0$ \cite{boesten}. 
For the  dipolar  $^{164}$Dy atoms, we take the intraspecies scattering length as
$a_1\equiv a(^{164}$Dy)$ =85a_0$ for the study of the symbiotic supersolid of the first type.
The yet unknown interspecies scattering length  is taken, after a small experimentation,  as    $a_{12}\equiv a(^{164}$Dy-$^{87}$Rb$)=-20a_0$ for the study of the symbiotic supersolid of the first type.  A larger value of the scattering length $a_{12}$ ($> -20a_0$) makes the system less attractive and consequently, leads to a   visible atom cloud of  $^{87}$Rb atoms, specially, for a larger number of  $^{87}$Rb atoms ($N_2\gtrapprox 10000$),  while the number  of  $^{164}$Dy atoms will be kept fixed at $N_1=50000$.

\begin{figure}[!t]

\begin{center}

\includegraphics[width=\linewidth]{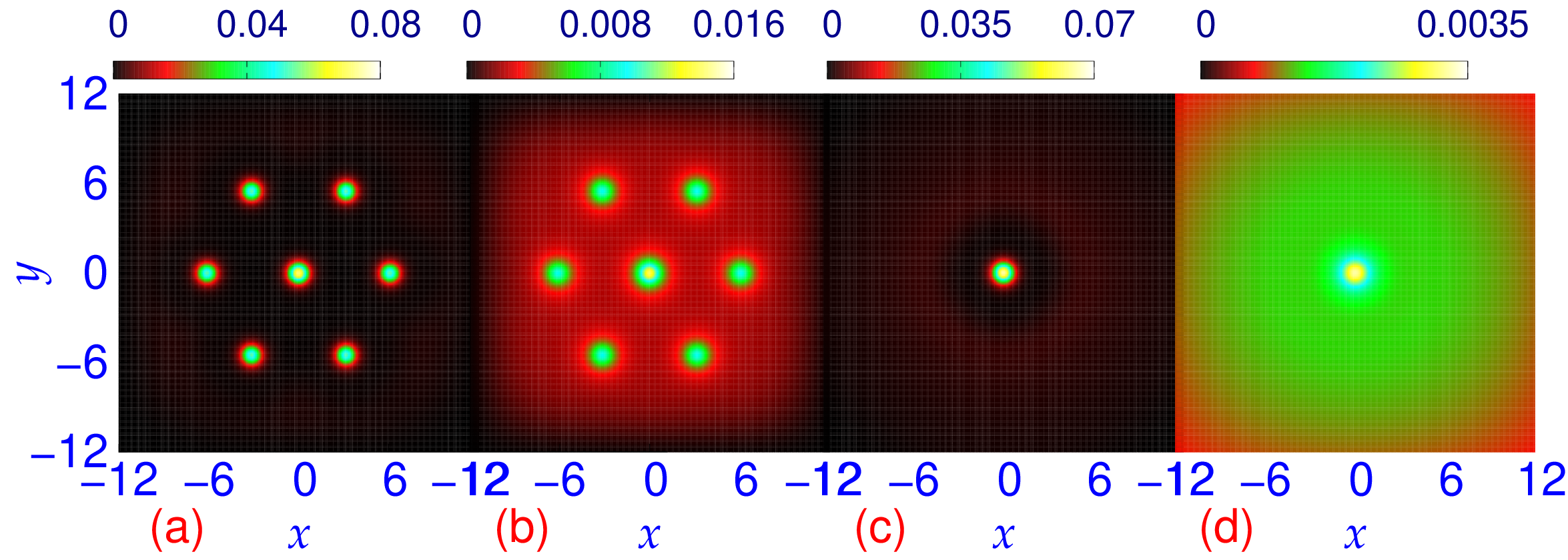}

\caption{ (Color online) Contour plot of quasi-2D density $n_1(x,y)$ and  $n_2(x,y)$, respectively,  of   seven droplets of 
(a) $^{164}$Dy and (b) 
$^{87}$Rb atoms for $a_{12}\equiv a(^{164}$Dy-$^{87}$Rb) $ =-5a_0$ 
in a 
  binary  $^{164}$Dy-$^{87}$Rb  mixture, where the $^{164}$Dy ($^{87}$Rb)   atoms are trapped   (untrapped). The same  for  $a_{12}=-2a_0$ are displayed in (c) and (d), respectively, for   $^{164}$Dy and $^{87}$Rb atoms.
  The number of atoms are
  $N_1\equiv  N(^{164}$Dy)$=50000$,   $N_2\equiv   N(^{87}$Rb)$  =10000$ in this figure and in Figs. \ref{fig5}, \ref{fig6}, \ref{fig8}, \ref{fig9}, and \ref{fig10}.
  }\label{fig4}
\end{center}
\end{figure}

We take $a_1 = 85a_0$  $(\ll a_{\mathrm{dd}} = 130.8a_0)$  to make the dipolar system 
strongly dipolar  with a relative dipolar length $\epsilon_{\mathrm{dd}}=1.539$, so that 
  a  droplet-lattice supersolid is easily formed;  even in the absence of an interspecies 
contact interaction, a dipolar supersolid  will  exist. 
We now study the formation of a quasi-free nondipolar $^{87}$Rb supersolid in a $^{164}$Dy-$^{87}$Rb mixture, where the dipolar  dysprosium atoms are trapped and the nondipolar rubidium atoms are untrapped.  After a small experimentation we fix $a_{12} =-20a_0$,
which is adequate for an efficient formation of a nondipolar $^{87}$Rb supersolid. Due to the  interspecies attraction the rubidium atoms stick to the dysprosium droplets and form a quasi-free supersolid pattern of nondipolar droplets.  If $|a_{12}|$ is taken smaller,  some rubidium atoms leave  the droplet pattern and form a background atom cloud.  With further reduced  $|a_{12}|$,  the supersolid  droplets of rubidium atoms will be lost creating a dense background atom cloud, while the supersolid dysprosium droplets remain intact. This is illustrated  in Fig. \ref{fig4}, 
for  a binary $^{164}$Dy-$^{87}$Rb
supersolid with seven droplets, obtained by imaginary-time propagation employing   identical seven-droplet state 
as the initial function in both components.
 We display in Fig. \ref{fig4} a contour plot of quasi-2D density  
of     (a)  $^{164}$Dy   and  (b) $^{87}$Rb atoms for $a_{12}=-5a_0$ and of  (c)  $^{164}$Dy  and  (d) $^{87}$Rb atoms for $a_{12}=-2a_0$ in a quasi-2D $^{164}$Dy-$^{87}$Rb mixture. The background atom cloud density is  already large in the  $^{87}$Rb supersolid in Fig.  \ref{fig4}(b) for  $a_{12}=-5a_0$
and in   Fig.  \ref{fig4}(d) the supersolid droplets have completely  disappeared in the background atom cloud
for $a_{12}=-2a_0$.

\begin{figure}[!t]
\begin{center}

\includegraphics[width=\linewidth]{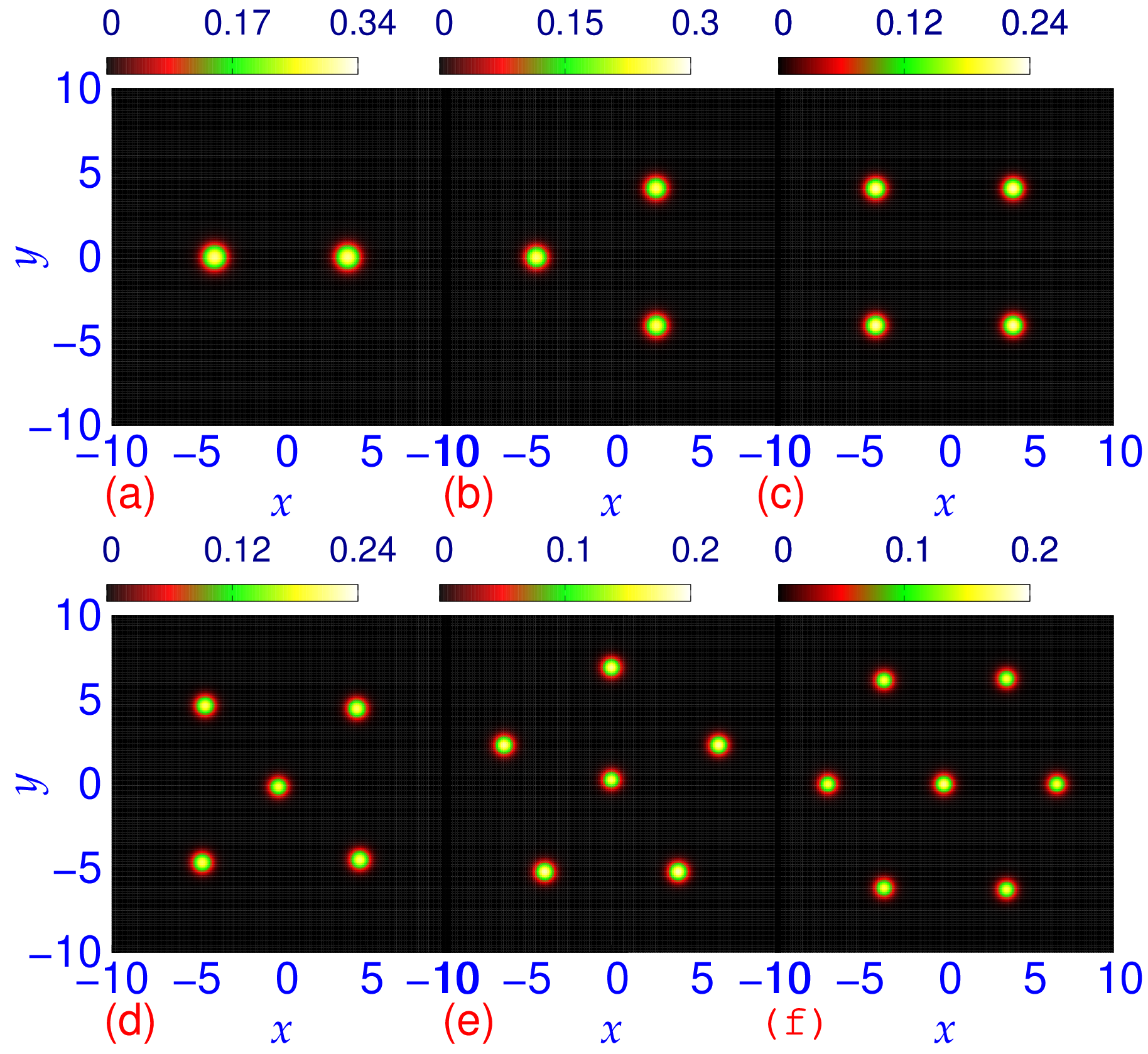}

\caption{ (Color online) Contour plot of quasi-2D density  $n_2(x,y)$ of (a) two, (b) three, (c) four,
(d) five, (e) six, and (f) seven droplets of 
$^{87}$Rb atoms in a 
  binary  $^{164}$Dy-$^{87}$Rb  mixture, where the $^{164}$Dy are trapped and the $^{87}$Rb atoms untrapped.  The interspecies scattering length is $a_{12}=-20a_0$. Other parameters are the same as in Fig. \ref{fig4}. 
  }\label{fig5}
\end{center}
\end{figure}

 In Fig. \ref{fig5} we illustrate the supersolid states of $N_1=50000$ dysprosium atoms and  $N_2=10000$ rubidium atoms for an interspecies scattering length $a_{12}=-20a_0$, through a contour plot of quasi-2D density $n_2(x,y)$ of $^{87}$Rb atoms,  
in the $^{164}$Dy-$^{87}$Rb mixture  with 
(a) two, (b) three, (c)  four, (d)  five, (e) six, and (f) seven droplets, respectively. These states were obtained by imaginary-time propagation employing an identical number of droplets in both components as the initial state.  Essentially, we consider   the supersolid droplet states of  Figs. \ref{fig2}(b)-(g)  
forming  two, three, four, five, six, and seven droplets, respectively, and  a small number  of untrapped   $^{87}$Rb atoms. 
  In all cases the numbers of overlapping droplets in $^{164}$Dy atoms and $^{87}$Rb atoms are the same.
   If we compare Figs. \ref{fig4} and \ref{fig5}, we see that the background atom cloud in $^{87}$Rb atoms is highly  reduced in Fig. \ref{fig5} as the interspecies scattering length $a_{12}$ has changed from $-2a_0$ to $-20a_0$ turning the binary dipolar-nondipolar mixture more attractive $-$ the  $^{87}$Rb atoms in the atom cloud have moved towards the center of the droplets, thus turning the  $^{87}$Rb  droplets more pronounced.

\begin{figure}[!t]

\begin{center}

\includegraphics[width=\linewidth]{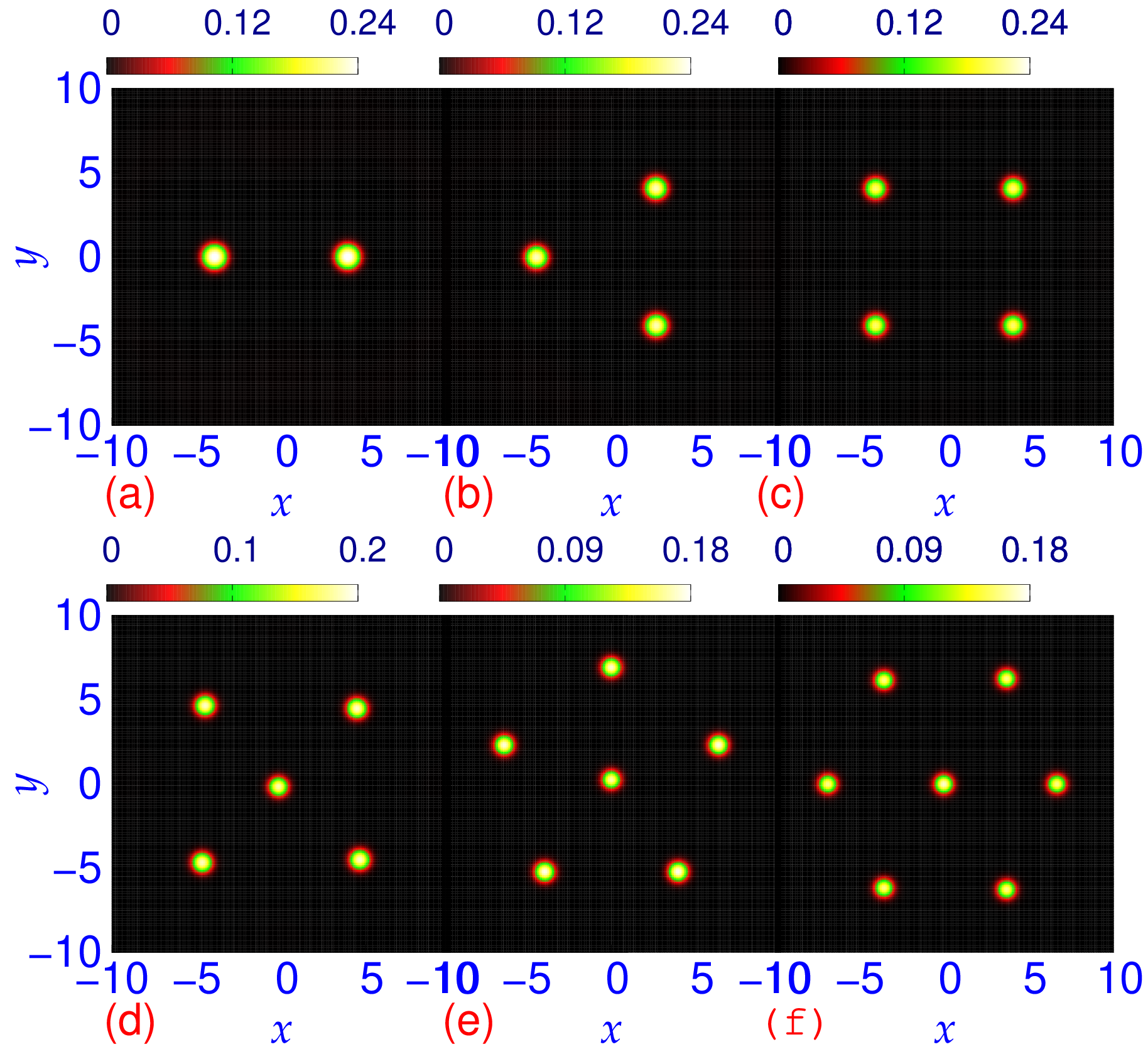}

\caption{ (Color online) Contour plot of quasi-2D  density  $n_1(x,y)$ of (a) two, (b) three, (c) four,
(d) five, (e) six, and (f) seven droplets of 
$^{164}$Dy atoms in the 
dipolar-nondipolarbinary  $^{164}$Dy-$^{87}$Rb  mixture of Fig. \ref{fig5}.  All parameters are the same as in   Fig. \ref{fig5}. 
  }
  \label{fig6}
\end{center}
\end{figure}

The quasi-2D density of  the   $^{164}$Dy droplet-lattice states 
    in the symbiotic dipolar-nondipolar   $^{164}$Dy-$^{87}$Rb   supersolids of   Fig.  \ref{fig5} are considered next.
  In  Fig. \ref{fig6}  we display the quasi-2D density  $n_1(x,y)$ of  $^{164}$Dy atoms in the dipolar-nondipolar mixture  of Figs. \ref{fig5} for (a) two, (b) three, (c)  four, (d)  five, (e) six, and (f) seven  droplets.
  The density profile of Figs. \ref{fig5}  and \ref{fig6} are similar qualitatively, although the maximum density for      $^{87}$Rb and  $^{164}$Dy atoms are quantitatively different. As expected, the droplets in the dipolar and nondipolar components are overlapping. However, if we compare the droplets  of the  $^{164}$Dy atoms in the  $^{164}$Dy-$^{87}$Rb   supersolid mixture, viz. Figs. \ref{fig6}(a)-(f), with  the same    in a single-component supersolid, viz. Figs. \ref{fig2}(b)-(g), we find that the supersolid droplets in the mixture are sharper and of high density,  essentially without any visible background atom cloud, whereas the same in the single-component case are wider, of low density, with a  reasonable amount of background atom cloud.   
This is because the attractive interspecies interaction has increased the binding of  the supersolid mixture 
so as to form a more tightly bound supersolid with sharper  droplets  in both components.  Hence an attractive interspecies interaction aids in forming sharper droplets. In Sec. 3.3  we will see that an attractive interspecies interaction can also transform a superfluid dipolar-nondipolar mixture without any droplet in both components into a  dipolar-nondipolar symbiotic   supersolid, where both components have identical droplet-lattice structure.

\begin{figure}[!t] 

\begin{center}
\includegraphics[width=\linewidth]{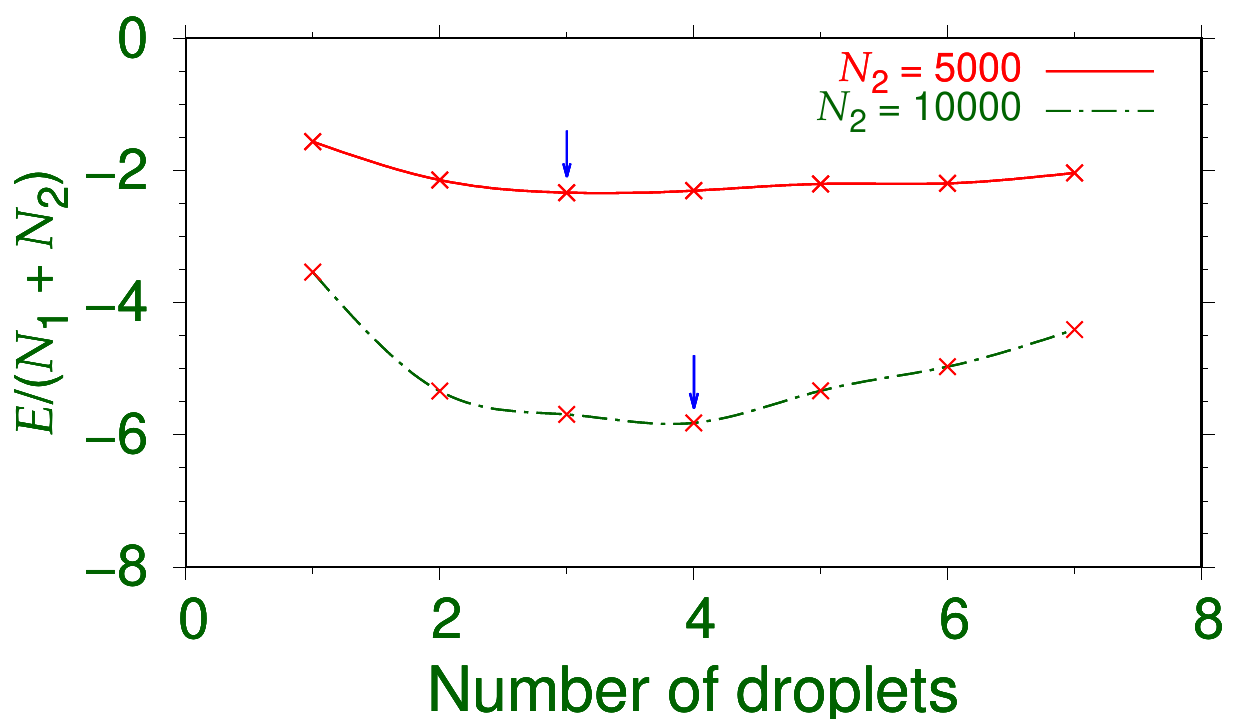}

\caption{ (Color online) Energy per atom $E/(N_1+N_2)$  of a dipolar-nondipolar $^{164}$Dy-$^{87}$Rb mixture,   for different number of spatially-symmetric droplets, as in Figs. \ref{fig4}-\ref{fig6}, of   $N_2$ $(= 5000,10000)$  $^{87}$Rb   atoms. The points are results of calculations and 
the lines indicate the trend of energy evolution.    The arrows indicate the minima of energy.  
Other parameters are the same as in Fig. \ref{fig5}.
 }\label{fig7}
\end{center}

\end{figure}

We next consider the  evolution of energy per atom $E/(N_1+N_2)$ of the $^{164}$Dy-$^{87}$Rb symbiotic  supersolid  mixture in Fig. \ref{fig7} for different number $N_2$ $(= 5000, 10000)$  of $^{87}$Rb atoms as the number of droplets increase, while the number of $^{164}$Dy atoms is    $N_1= 50000$.  An interesting feature of this plot is that the energy per atom has a minimum  as the number of droplets increase from one   to seven. For $N_2=5000$ the energy has a  minimum   for three droplets, whereas for 
 $N_2=10000$   the energy has a minimum for four droplets, both indicated by an arrow in Fig. \ref{fig7}. Hence, in an experiment, from an energetic consideration, 
 the three-droplet state is more likely for  $N_2=5000$  and the four-droplet state is more likely for 
 $N_2=10000.$  
 This is consistent with Fig. \ref{fig3}, where we find that in the single-component case,  the three-droplet state is most likely from an energetic consideration.  The addition of a small number of   $N_2=5000$    $^{87}$Rb atoms  to $N_1=50000 (\ll N_2)$ $^{164}$Dy atoms is a small perturbation and the three-droplet state continues to be more probable.  But the addition of   $N_2=10000$   $^{87}$Rb atoms corresponds to a larger change in energy and makes the system more attractive. Consequently, the most probable state is the four-droplet state. 
  In  order to have a supersolid state with a much  larger number of droplets, both  the number of 
dysprosium atoms $N_1$ and the number of   rubidium atoms $N_2$ should be increased.

 \subsection{Symbiotic dipolar-nondipolar supersolid of the second type}
 
 For the study of a symbiotic supersolid of the second type, a larger value of the scattering length $a_1$ and a smaller  $a_{12}$ will be needed, 
which can be achieved experimentally by the Feshbach resonance technique \cite{fesh}.
Consequently, we take $a_1\equiv a(^{164}$Dy) $=95a_0$
to make the dipolar system weakly dipolar, so that no dipolar droplet-lattice supersolid appears without any interspecies attraction
with zero interspecies scattering length $a_{12}\equiv a(^{164}$Dy-$^{87}$Rb)=0.  With an adequate interspecies attraction both components 
develop overlapping supersolid droplet-lattice structure.  This is an ideal superfluid to supersolid transition, where both components become supersolid simultaneously from a superfluid phase. In the symbiotic supersolid  of the first type, the dipolar component is a supersolid even in the absence of interspecies attraction ($a_{12}=0$). The superfluid to supersolid transition in a symbiotic supersolid of the second type
 is a different type of transition, which is caused  by an increase in interspecies attraction and  not by a reduction in intraspecies contact 
repulsion  to make the binary system more attractive as in the case of a single-component dipolar superfluid to supersolid transition or of a symbiotic supersolid of the first type \cite{drop1,drop2}.

\begin{figure}[!t]

\begin{center}

\includegraphics[width=\linewidth]{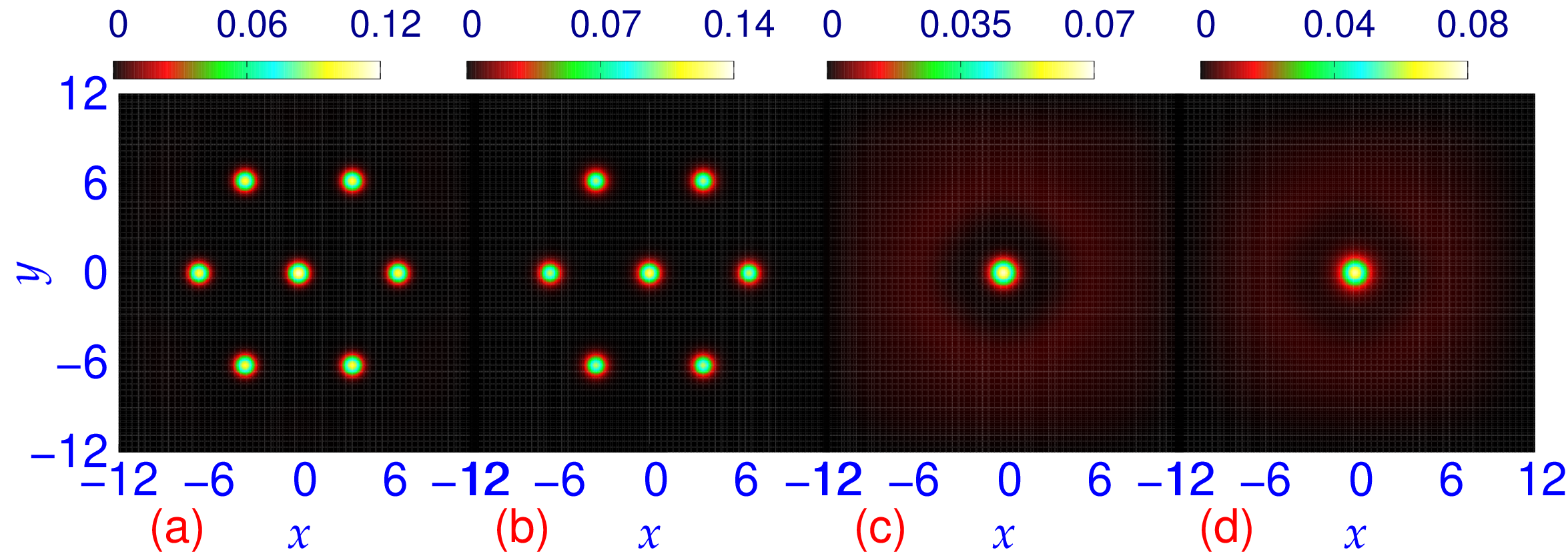}

\caption{ (Color online) Contour plot of quasi-2D density  $n_1(x,y)$  and  $n_2(x,y)$, respectively,  of   seven droplets of 
(a) $^{164}$Dy atoms and (b) 
$^{87}$Rb atoms for  
 $a_1\equiv  a(^{164}$Dy)  $=90a_0$
in a 
  binary  $^{164}$Dy-$^{87}$Rb  mixture, where the $^{164}$Dy are trapped and the $^{87}$Rb atoms untrapped. The same  for $a_1  =95a_0$  are displayed in (c) and (d), respectively, for   $^{164}$Dy and $^{87}$Rb atoms.
Other parameters are the same as in Fig. \ref{fig5}.
  }\label{fig8}
\end{center}
\end{figure}

To demonstrate the effect of increasing the scattering length $a_1$ of dipolar atoms, we display in Figs. \ref{fig8}(a)-(b) contour plots of quasi-2D densities of (a) $^{164}$Dy and (b) $^{87}$Rb atoms for $a_{12}=-20a_0$, $a_1=90a_0$, $N_1=50000, N_2=10000$, where only the dysprosium atoms are trapped.   Although, a clean droplet-lattice supersolid emerges in both components in this case in the presence of an interspecies attraction ($a_{12}=-20a_0$), for zero interspecies scattering length  ($a_{12}=0$) the quasi-2D  density of dysprosium atoms have an extended  Gaussian profile without any droplets.  In Figs. \ref{fig8}(c)-(d)
contour plots of quasi-2D densities of (c) $^{164}$Dy and (d) $^{87}$Rb atoms, for $a_{12}=-20a_0$, $a_1=95a_0$, $N_1=50000,
N_2=10000$, are shown. With increased repulsion due to an augmentation of $a_1$, only a single droplet appears in both components.    A droplet-lattice structure only appears when the binary dipolar-nondipolar system is sufficiently attractive and this happens for a reduced scattering length of dipolar atoms, viz. Figs. \ref{fig6}-\ref{fig7}, or for an increased interspecies attraction,  as demonstrated in the following.

\begin{figure}[!t]

\begin{center}

\includegraphics[width=\linewidth]{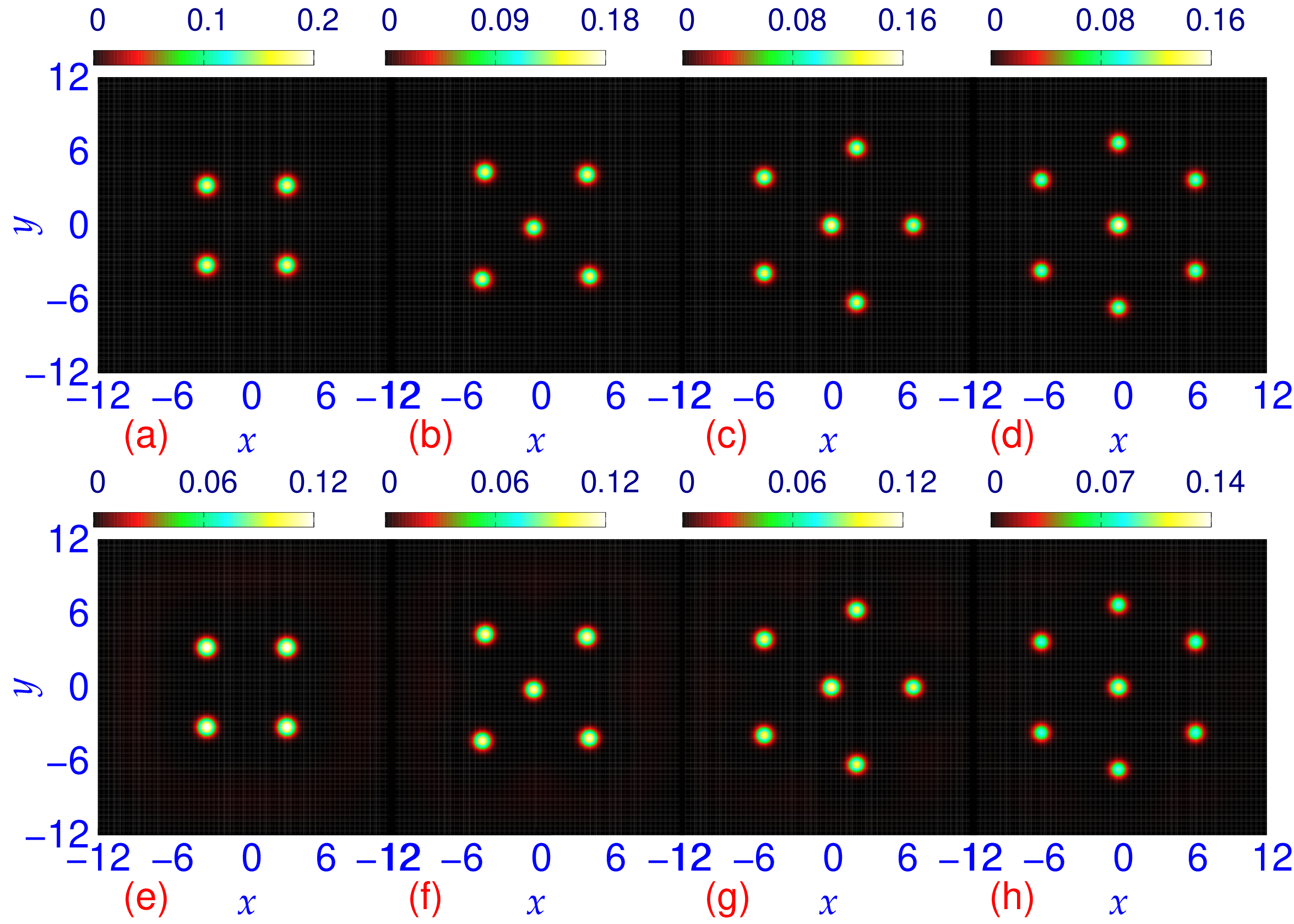}

\caption{ (Color online) Contour plot of quasi-2D density    of 
(a) four, (b) five,  (c) six,  (d) seven droplets of  
$^{87}$Rb atoms in a 
   binary  $^{164}$Dy-$^{87}$Rb  mixture, 
where only  $^{164}$Dy atoms are trapped. 
The same    of 
(e) four, (f) five,  (g) six,   (h)  seven droplets of 
$^{164}$Dy atoms in the same binary   mixture. 
Other parameters are  $a_1 = 95a_0, a_2= 109a_0, a_{12}
   =-25a_0$ in this figure and  in Fig. \ref{fig10}.    }
  \label{fig9}
\end{center}
\end{figure}

 To illustrate the realization of a supersolid of the second type we  consider a binary 
 $^{164}$Dy-$^{87}$Rb mixture with $a_1=95a_0, a_2=109a_0, a_{12}=-25a_0$ and $N_1=50000, N_2=10000$.
 In this case, for a relatively weaker dipolar attraction with a relative dipolar length 
 $\epsilon_{\mathrm{dd}}=1.377$, compared to $\epsilon_{\mathrm{dd}}=1.539$ in Sec. 3.2 for the supersolid of the first type, there are no droplets in  the dipolar atoms for $ a_{12}=0.$  However, for an adequate interspecies attraction $a_{12}=-25a_0$,
a  droplet-lattice supersolid appears, simultaneously, in both the dipolar $^{164}$Dy and nondipolar $^{87}$Rb atoms. In Fig. \ref{fig9} we display a contour plot of quasi-2D density $n_2(x,y)$ of (a) four, (b) five, 
(c) six, and (d) seven droplets of the $^{87}$Rb atoms in the binary dipolar-nondipolar mixture, where only the 
dipolar atoms are trapped. In Figs. \ref{fig9}(e)-(h) we illustrate the corresponding densities of the  
 $^{164}$Dy stoms.  The droplets in the two components in Figs. \ref{fig9}  are quite similar and overlapping.

 \subsection{Generation of a  dipolar-nondipolar supersolid in a laboratory}

\begin{figure}[!t]

\begin{center}

\includegraphics[width=\linewidth]{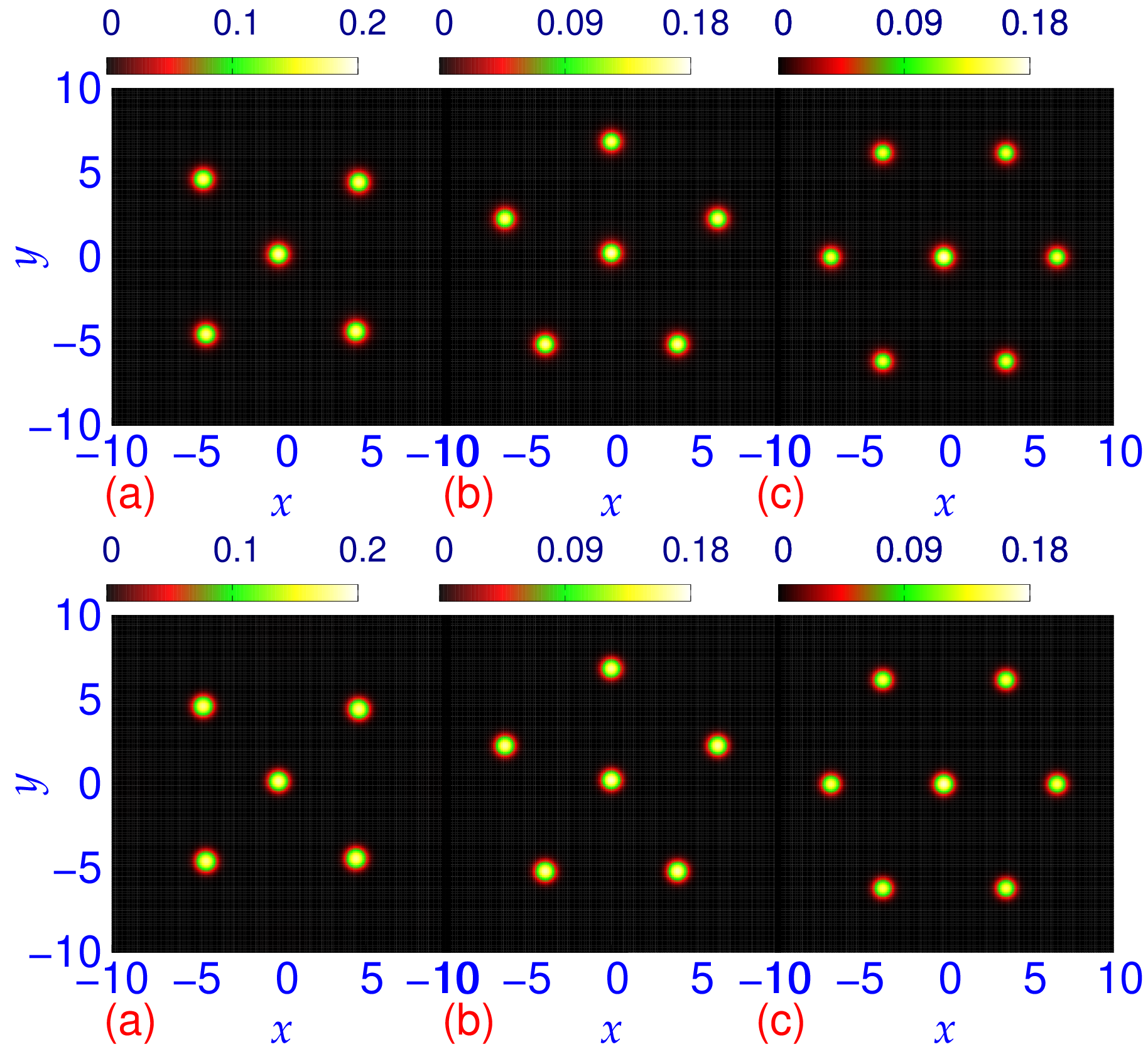}

\caption{ (Color online) Contour plot of quasi-2D density  of 
(a) five, (b) six, and (c) seven droplets of  
$^{87}$Rb atoms in the  binary  $^{164}$Dy-$^{87}$Rb  mixture, where both components are trapped. 
The same  of 
(a) five, (b) six, and (c) seven droplets of 
$^{164}$Dy atoms in the same  mixture. 
  }
  \label{fig10}
\end{center}
\end{figure}

The present quasi-free nondipolar droplet-lattice supersolid  is not just of theoretical interest, but can  be realized experimentally by initially preparing a binary $^{164}$Dy-$^{87}$Rb  overlapping supersolid mixture where both the components are  harmonically 
trapped. The trap on $^{87}$Rb can then be  ramped to zero to find out if the desired quasi-free nondipolar $^{87}$Rb supersolid state can be formed.
  To illustrate numerically 
the viability of this procedure,  we consider 
an initial  $^{164}$Dy-$^{87}$Rb binary supersolid mixture with five, six and seven droplets for the following parameters
$N_1 = 50000, N_2  = 10000,$ $a_1=85a_0, a_2=109a_0, a_{12}=-20a_0$.    We   take the same harmonic trap acting on both components. The quasi-2D density profile of    $^{87}$Rb  and   $^{164}$Dy  atoms,  calculated by imaginary-time propagation,
for five, six and seven droplets, are displayed in Figs. \ref{fig10}(a)-(c)   and     Figs. \ref{fig10}(d)-(f), respectively.  
We find that these densities in Figs. \ref{fig10}(a)-(c)   and     Figs. \ref{fig10}(d)-(f) are quite similar to the densities in Figs. \ref{fig5}(d)-(f)   and     Figs. \ref{fig6}(d)-(f), for five, six, and seven droplets, of $^{87}$Rb  and   $^{164}$Dy  atoms
respectively. The  densities in Figs. \ref{fig5}(d)-(f)   and     Figs. \ref{fig6}(d)-(f) refer to the case where only the 
 $^{164}$Dy  atoms are trapped and the  $^{87}$Rb    atoms are untrapped.  The harmonic trap on the rubidium atoms has negligible 
 effect on the density of the traps, as these atoms are essentially confined by the interspecies attraction and  not by 
 the harmonic trap. But the trap on the rubidium atoms is essential to experimentally localize the  $^{87}$Rb atoms in the form of a supersolid droplet state.  Once such a state is formed, the trap on the   $^{87}$Rb atoms can be removed 
to prepare a robust quasi-free supersolid state of   rubidium atoms. In fact we performed a real-time propagation employing the 
fully trapped states of Fig. \ref{fig10} as the intial state, after removing the trap on the rubidium atoms.
We find that  the system quickly produces  the states of   Figs. \ref{fig5}(d)-(f)   and     Figs. \ref{fig6}(d)-(f), establishing the claim of realizing the symbiotic droplet-lattice supersolid  in a laboratory. We do not explicitly show the evolution of the density pattern in this numerical experiment as there is practically no visible change of density profiles of the states.


\section{Summary and Discussion}

Using a numerical solution   of a   set of improved binary mean-field GP equations, 
we demonstrate the existence   of two types of quasi-2D symbiotic supersolids in a  dipolar-nondipolar 
$^{164}$Dy-$^{87}$Rb mixture by imaginary-time propagation, where the dipolar atoms are trapped and nondipolar atoms untrapped, in the presence of an adequately attractive interspecies interaction.
In the absence of interspecies interaction [uncoupled case, $a_{12}\equiv a(^{164}$Dy-$^{87}$Rb) =0], in the 
first type the dipolar atoms are in a droplet-lattice supersolid state, whereas in the second type the dipolar atoms are in a superfluid  state without any droplet. In both cases  an untrapped quasi-free  nondipolar $^{87}$Rb droplet-lattice supersolid, 
stabilized by an interspecies attraction, is formed.  In the first type of symbiotic supersolid, to have a droplet-lattice supersolid 
in $^{164}$Dy atoms  for $a_{12}=0$, a strongly dipolar system is needed; in the present study we considered 
$a_1 \equiv a(^{164}$Dy) $= 85a_0$, corresponding to the relative dipolar length $\epsilon_{\mathrm{dd}}= 1.539$ and $a_{12} \equiv a(^{164}$Dy-$^{87}$Rb) $= -20a_0$.     In the second type of symbiotic supersolid, there are no droplets  
in $^{164}$Dy atoms  for $a_{12}=0$ and to achieve this   a weaker  dipolar system is necessarily needed. This is achieved in this study   employing a larger value of  the scattering length of dipolar atoms 
$a_1 = 95a_0$, corresponding to the relative dipolar length $\epsilon_{\mathrm{dd}}= 1.377$ and $a_{12} = -25a_0$, compared to a larger value of   the relative dipolar length $\epsilon_{\mathrm{dd}}= 1.539$ in Sec. 3.1 for the first type of symbiotic supersolid.

 We also demonstrate numerically 
that  a symbiotic droplet-lattice   supersolid  can be obtained experimentally by 
considering a  binary $^{164}$Dy-$^{87}$Rb BEC, where both components are trapped, and then removing the trap on $^{87}$Rb. In this numerical study we prepare the   binary $^{164}$Dy-$^{87}$Rb BEC, where both components are trapped, by imaginary-time propagation. Then we perform a real-time propagation after a removal of the trap on $^{87}$Rb atoms to obtain the desired symbiotic droplet-lattice   supersolid. 
As the  density profiles of the binary mixture with the $^{87}$Rb atoms trapped and untrapped are 
practically the same, hence, after removal of the trap on $^{87}$Rb atoms,  the desired symbiotic supersolid is quickly obtained.

\section*{CRediT authorship contribution statement}

S. K. Adhikari: Conceptualization, Methodology, Validation, Investigation, Writing – original draft, Writing – review and  editing, Supervision, Funding aquisition, 
Visualization.

\section*{Declaration of competing interest}
The authors declare that they have no known competing financial interests or personal relationships that could have appeared
to influence the work reported in this paper.

\section*{Data availability}
No data was used for the research described in the article.

\section*{Acknowledgments}

SKA acknowledges support by the CNPq (Brazil) grant 301324/2019-0.

\end{document}